\documentstyle[aps,floats]{revtex}
\input epsf.tex
\input amssym.def
\def\half{\textstyle{1\over 2}}

\begin{document}
\title{Restoration of particle number as a good quantum number in BCS theory}
\author{D.J.\ Rowe}
\address{ Department of Physics, University of Toronto,
Toronto, Ontario M5S 1A7, Canada}
\date{Aug, 2000}
\maketitle
\begin{abstract} 
As shown in previous work, number projection can be carried out analytically for
states defined in a quasi--particle scheme when the states are expressed in a
coherent state representation.
The wave functions of number--projected states are
well--known in the theory of orthogonal polynomials as Schur functions.
Moreover, the functions needed in pairing theory are a particularly simple class
of Schur functions that are easily constructed by means of recursion relations.
It is shown that complete sets of states can be projected from corresponding
quasi--particle states and that such states retain many of the properties of the
quasi--particle states from which they derive.
It is also shown that number projection can be used to construct a complete
set of orthogonal states classified by generalized seniority for any
nucleus.
\end{abstract}
\pacs{02.20.-a, 03.65.Fd, 21.60.Ev}

\section{Introduction}

The loss of particle number as a good quantum number is a serious deficiency of
the BCS and HFB (Hartree--Fock--Bogolyubov) approximations in their applications to
finite nuclei.
However, it can be restored with relative ease in a
coherent-state representation \cite{RSC,CSR} in a manner that preserves much of
the elegance and simplicity of the models.

Following early suggestions by Mottelson \cite{Mottelson}, number conserving
extensions of BCS and HFB theory have been considered by many authors: cf., for
example, the number-projected BCS approximation of Dietrich, Mang, and Pradal
\cite{DMP}, the projected quasi-particle model of Lande, Ottaviani and Savoia
\cite{projqp}, the antisymmetrized geminal power states of Coleman {\em et al.}
\cite{antigem}, the broken pair model of Lorazo, Gambhir, and others
\cite{Lorazo,Gambhir}, Talmi's generalized seniority scheme \cite{Talmi}, and the
coherent correlated pair method of  Vary and Plastino \cite{VP}.
A review of the subject has been given by Allaart {\em et al.}
\cite{BPA}.

A common feature of these schemes is to approximate the ground state, for a
system in which pair correlations dominate, by a state of the form
\begin{equation} |{\cal N}\rangle = \Pi_{N} |x\rangle \,,\end{equation}
where $|x\rangle$ is a quasi-particle vacuum state and $\Pi_{N}$
is a projector onto the space of states of particle number $N$.
We show herein that such a projection can be applied analytically, when the wave
functions are expressed in a coherent-state representation.
More importantly, projection from quasi-particle states gives a complete set of
$N$-particle states which reflect many of the properties of the states from
which they were projected.
Thus, a potentially exact formalism emerges which is able to take advantage of the
physical insights gained from the BCS and HFB approximations.

Analytic number projection when combined with analytic methods for
projecting out states of good angular momentum from deformed intrinsic states
\cite{AMproj} raises the possibility of developing good theories of nuclei which
exhibit both rotational and superconducting properties.
It also suggests a way of extending the physically
significant definition of generalized seniority to the classification of complete
sets of orthonormal states.

For simplicity, we restrict consideration in this paper to the BCS 
model.
However, the methods extend to the general HB theory by the number projection
methods given in ref.\ \cite{RSC}.

\section{Quasispin algebras and their coherent state representation} 

For each $j$-shell, there is an SU(2)  quasispin algebra spanned by the operators
\begin{eqnarray} \displaystyle &\hat S^j_+=\sum_{m=1/2}^{j} 
a^\dagger_{jm} a^\dagger_{j\bar m},\quad \hat S^j_- =
\sum_{m=1/2}^{j}a^{j\bar m} a^{j m} \, ,&\nonumber\\
&\hat S^j_0 = \displaystyle {1\over 4}\sum_{m=-j}^{j} (
a^\dagger_{jm} a^{jm} - a^{jm}a^\dagger_{jm})
\,,&\end{eqnarray}
where 
\begin{equation} a^\dagger_{j\bar m} =  (-1)^{j-m} a^\dagger_{j,-m} \,,\quad
a^{j\bar m} = (-1)^{j-m} a^{j,-m} = a_{jm} \,,\end{equation}
are creation and annihilation
operators for a nucleon (e.g., neutron) which satisfy the fermion
anticommutation relations
\begin{equation} \{ a^{jm}, a^\dagger_{jn}\} = \delta_{mn} \,.\end{equation}
Note that we follow Einstein's convention of using upper indices to label the
components of a tensor contragredient to a tensor labeled by lower indices.
With this convention, the annihilation operator $a^{jm}$ with superscripts is the
Hermitian adjoint of the creation operator $a^\dagger_{jm}$ and the annihilation
operator $a_{jm}$ with subscripts is the
$m$ component of a spherical tensor.
The contraction of a pair of upper and lower indices is then a scalar; e.g.,
\begin{equation} \hat n_j =\sum_m a^\dagger_{jm} a^{jm}  \end{equation}
is the number operator for the $j$-shell.
The quasispin operators satisfy the usual SU(2) commutation
relations
\begin{equation} [\hat S^j_+, \hat S^j_-] = 2\hat S^j_0 \, ,\quad
[\hat S^j_0, \hat S^j_\pm] = \pm  \hat S^j_\pm \, ,\end{equation}
and the nucleon operators transform under SU(2) as components of
quasispin--$1/2$ tensors:
\begin{equation} \begin{array}{rcl}
[\hat S^j_-, a^\dagger_{jm}] = a_{jm},&\quad &[\hat S^j_+\,,
a_{jm}]=a^\dagger_{jm}\,, 
\cr [\hat S^j_0, a^\dagger_{jm}] = \half a^\dagger_{jm}\,,&& 
[\hat S^j_0, a_{jm}]=- \half a_{jm}\,.  
\cr\end{array}
\end{equation}

Representations of a quasispin algebra are characterized by highest and lowest
weights. For example, the zero-particle state $|0\rangle$ is a lowest weight
state for all the quasispin algebras;
\begin{equation} \hat S^j_-|0\rangle = 0 \,,\quad \hat S^j_0|0\rangle = -s_j |0\rangle \,,
\quad \forall\, j \,,\end{equation} 
where $s_j = (2j+1)/4$.
Such a state is said to have multishell seniority $(0,0,\ldots)$.
More generally, a state $|\nu JM\rangle$ of $\nu$ particles that satisfies the
equations
\begin{equation} \hat S^j_-|\nu JM\rangle = 0 \,,\quad \hat S^j_0|\nu JM\rangle =
(-s_j+\nu_j/2) |\nu JM\rangle \,,\quad \forall\, j \,,\end{equation}
is said to have seniority $\nu $ and multishell seniority
$(\nu_{j_1},\nu_{j_2},
\ldots )$, where $\nu = \sum_j \nu_j$.
The multishell seniority defines the values of the lowest weights and provides
useful identifiers of the corresponding quasispin irreps.

A coherent state for the combined quasispin algebras is defined by
\begin{equation} |x\nu\rho JM\rangle = e^{\hat S_+(x)} |\nu\rho JM\rangle \quad 
{\rm with}
\quad
\hat S_+(x) =
\sum_j x_j \hat S^j_+\,,\end{equation}
where $|\nu\rho JM\rangle$ is a lowest weight state and $\rho$ provides whatever
labels are needed, in addition to $\nu JM$, to characterize the state.
 A coherent state is a quasispin transform of a lowest weight state.
The simplicity of the BCS and HFB theories stems from the fact that the
quasi-particle operators of the Bogolyubov--Valatin transformation are also
quasispin transforms of  nucleon creation and annihilation operators.
In particular,  a coherent state of the seniority-zero irrep 
\begin{equation} |x\rangle = e^{\hat S_+(x)} |0\rangle \end{equation}
is the vacuum state of the quasi-particles defined, for real values of $\{x_j\}$,
by the 
 Bogolyubov--Valatin transformation
\begin{equation} \begin{array}{c}\alpha^\dagger_{jm} =  u_j  a^\dagger_{jm}- v_j a_{jm} \,,\\
\alpha_{jm} =  u_j  a_{jm}+ v_j a^\dagger_{jm} \,.
\end{array}\end{equation}
with $u_j^2+v_j^2=1$ and $v_j = x_j u_j$.
This well-known result follows from the observation that 
\begin{equation} \alpha_{jm}  e^{\hat S_+(x)} =  e^{\hat S_+(x)} (u_j a_{jm}+v_j a^\dagger_{jm}
-x_ju_j a^\dagger_{jm}) =  u_j  e^{\hat S_+(x)} a_{jm} \,.
\end{equation} 

Although the transformation $e^{\hat S_+(x)}$ is not unitary, it is
simply related to the unitary transformation 
\begin{equation} g(x) = e^{\hat S_+(\beta) -\hat S_-(\beta)} \,, \end{equation}
for which 
\begin{equation} \cos \beta_j = {1\over \sqrt{1+x_j^2}} = u_j \,,\quad \sin \beta_j = {x_j\over
\sqrt{1+x_j^2}} = v_j \,.\end{equation}
The relationship is given by the expansion 
\begin{equation} g(x) = e^{\hat S_+(x)} \prod_j(1+x_j^2)^{\hat S^j_0}\, e^{-\hat S_-(x)}
\,.\end{equation} 
Thus,  the transformation
\begin{equation} |0\rangle \to |x\rangle = e^{\hat S_+(x)}|0\rangle = g(x) |0\rangle \, \prod_j
(1+ x_j^2)^{s_j}\end{equation}
is unitary, to within a factor $\sqrt{\Phi^s(x^2)}$, where
\begin{equation} \Phi^s(x^2) = \langle x|x\rangle = \prod_j (1+ x_j^2)^{2s_j}  \end{equation}
and $x^2 =(x_{j_1}^2, x^2_{j_2}, \ldots )$.
Likewise, the BV transformation is seen as the unitary SU(2)
transformation
\begin{equation} a^\dagger_{jm} \to \alpha^\dagger_{jm} = g(x) a^\dagger_{jm} g^\dagger(x)
\,,\quad
 a_{jm} \to \alpha_{jm} = g(x) a_{jm} g^\dagger(x) \,.\end{equation}

In a VCS (vector coherent state) representation \cite{VCS,RR91}, a lowest
weight state $|\nu\rho JM\rangle$ is regarded as an intrinsic state for an irrep
of the  combined quasi-spin algebras and represented by an intrinsic wave function
$\xi_{\nu\rho JM}$.
An arbitrary state $|\psi\rangle$ is then represented by a holomorphic wave
function
$\Psi$ defined over a set of complex coordinates $z=(z_{j_1},z_{j_2},
\ldots )$  with values 
\begin{equation} \Psi(z) = \sum_{\nu\rho JM} \xi_{\nu\rho JM} \langle {\nu\rho JM}|e^{\hat
S_-(z)} |\psi\rangle \,.\end{equation}
For example, if the zero-particle vacuum $|0\rangle$ has
intrinsic wave function $\xi_0$, an arbitrary quasi-particle vacuum state
$|x\rangle=e^{\hat S_+(x)} |0\rangle$ has wave function $\Psi_x$ with values
\begin{equation} \Psi_x(z) = \xi_0 \langle 0 |e^{\hat S_-(z)}e^{\hat S_+(x)} |0\rangle  
= \xi_0 \Phi^s(zx)\,,\end{equation}
where $\Phi^s(zx)$ is the overlap
\begin{equation} \Phi^s(zx) = \langle z |x\rangle = \prod_j (1+z_jx_j)^{2s_j} \end{equation}
with $zx = ( z_{j_1}x_{j_1}, z_{j_2}x_{j_2}, \ldots )$.
It follows that the norm of a quasi-particle vacuum state $|x\rangle$ is
 the value of the function $\Phi^s$ at $x^2$.

In the VCS representation, an element $\hat X$ of an SU(2) quasispin algebra is
represented as a differential operator $\Gamma(\hat X)$ defined by 
\begin{eqnarray}  [\Gamma (\hat X)\Psi](z) &=&   \sum_{\nu\rho JM} \xi_{\nu\rho JM} \langle
{\nu\rho JM}| e^{\hat S_-(z)}\hat X |\psi\rangle  \nonumber\\
&=&  \sum_{\nu\rho JM} \xi_{\nu\rho JM} \langle
{\nu\rho JM}|\big(\hat X + [\hat
S_-(z),X] \nonumber\\ 
&&\quad + \half [\hat S_-(z),[\hat S_-(z),X]]\big) e^{\hat S_-(z)} |\psi\rangle \,.
\end{eqnarray}
Evaluation of the right hand side of this expression by means of the
identities
\begin{eqnarray} &\langle{\nu\rho JM}|\hat S_+^j  e^{\hat S_-(z)} |\psi\rangle = 0 \,,\quad
\langle{\nu\rho JM}|\hat S_0^j  e^{\hat S_-(z)} |\psi\rangle = (-s_j+\nu_j/2)
\Psi(z)
\,,& \nonumber\\
&\displaystyle\langle{\nu\rho JM}|\hat S_-^j  e^{\hat S_-(z)} |\psi\rangle =
{\partial\over
\partial z_j} \Psi(z) \,,&
\end{eqnarray}
gives the representation
\begin{equation} 
\Gamma (\hat S^j_0) = -\hat s_j+ z_j{\partial\over \partial z_j} \,,\quad
\Gamma (\hat S^j_-) ={\partial\over \partial z_j} \,,\quad
\Gamma (\hat S^j_+) = 
z_j\Big( 2\hat s_j - z_j{\partial\over \partial z_j}\Big) \,,
\end{equation}
where $\hat s_j$ is a diagonal operator on intrinsic wave functions with
eigenvalues
\begin{equation} \hat s_j \xi_{\nu\rho JM}= (-s_j+\nu_j/2) \xi_{\nu\rho JM} \,.\end{equation}
The number operator $\hat n_j$ is represented
\begin{equation} \Gamma (\hat n_j) = \hat \nu_j + 2z_j{\partial\over \partial z_j} \,.\end{equation}

\section{Number-projected states}

Let $|n(x)\rangle$ denote the $n$-pair state
\begin{equation} |n(x)\rangle = \big[ S_+(x)\big]^n |0\rangle
= n!\, \Pi_{2n}\, |x\rangle \,.\end{equation} 
States of good nucleon number can also be projected
from excited quasi-particle states.
Consider first the one quasi-particle state $\alpha^\dagger_{jm} |x\rangle$.
To within a factor $u_j= 1/\sqrt{1+x_j^2}$, the $(2n+1)$-particle component of
such a state is given by
\begin{equation}  u_j \Pi_{2n+1}  \alpha^\dagger_{jm} |x\rangle = \Pi_{2n+1}
e^{\hat S_+(x)} [u_j^2 + x_j u_jv_j] a^\dagger_{jm} |0\rangle =
{1\over n!} \big[ S_+(x)\big]^n |jm\rangle \,,\end{equation}
where $|jm\rangle = a^\dagger_{jm}|0\rangle$.
This is a special case of a general result. 

\medskip\noindent{\bf Claim 1:}  Let $Z_{\mu\beta JM}^\dagger (a^\dagger)$ denote
some combination of $\mu$ nucleon creation operators
\begin{equation} Z_{\mu\beta JM}^\dagger (a^\dagger) = \sum_{j} C^{j_1,\ldots ,j_\mu
}_{\mu\beta JM} [a^\dagger_{j_1}\otimes a^\dagger_{j_2}\otimes\cdots
\otimes a^\dagger_{j_\mu} ]_{JM}\end{equation}  which creates a state
\begin{equation} |\mu\beta JM\rangle =  Z_{\mu\beta JM}^\dagger (a^\dagger) |0\rangle\end{equation}
having the property that 
\begin{equation} \hat S_-(uv)|\mu\beta JM\rangle = 0 \,;\end{equation}
the subscript $\beta$ provides whatever labels are needed in addition to
$\mu JM$  to specify the state.
Let $Z_{\mu\beta JM}^\dagger (u\alpha^\dagger)$ denote the corresponding
quasi-particle operator in which each $a^\dagger_{jm}$ is replaced by
$u_j\alpha^\dagger_{jm}$.
Then
\begin{equation} \Pi_{2n+\mu} Z_{\mu\beta JM}^\dagger (u\alpha^\dagger) |x\rangle =     
{1\over n!} \big[ S_+(x)\big]^n |\mu\beta JM\rangle \,.\end{equation}

\medskip\noindent{\bf Proof:}  The claim follows from the identity
\begin{equation} u_j \alpha^\dagger_{jm} e^{\hat S_+(x)} =  e^{\hat S_+(x)} (a^\dagger_{jm} -
u_jv_j a_{jm}) =
 e^{\hat S_+(x)}  e^{-\hat S_-(uv)}a^\dagger_{jm} e^{\hat S_-(uv)} \end{equation}
which implies that
\begin{equation} Z_{\mu\beta JM}^\dagger (u\alpha^\dagger) |x\rangle =  e^{\hat S_+(x)} 
e^{-\hat S_-(uv)} Z_{\mu\beta JM}^\dagger (a^\dagger) |0\rangle = 
 e^{\hat S_+(x)} |{\mu\beta JM}\rangle \,.
\end{equation}
\hfil Q.E.D.
\medskip

It will be noted that to obtain complete sets of linearly independent states by
number projection in this way, the states $\{ |\mu\beta JM\rangle\}$ should
include not only states that satisfy
\begin{equation} \hat S^j_- |\mu\beta JM\rangle = 0 \quad \forall j\,, \label{eq:2.35}\end{equation}
but also states such as $\hat S_+(y) |\mu-2, \beta' JM\rangle$ for $y\not= x$.
Imposing the constraint
\begin{equation} \hat S_-(uv) \hat S_+(y) |\mu-2, \beta' JM\rangle =0 \end{equation}
for all such states is one way, but not the only way, of ensuring that no linear
combination of such states is of the form $\hat S_+(x) |\mu-2, \beta' JM\rangle$.
The latter states are not needed because the $(2n+\mu)$-particle states projected
from $e^{\hat S_+(x)}\hat S_+(x)|\mu-2, \beta' JM\rangle$ are proportional to
those projected from $e^{\hat S_+(x)}|\mu-2, \beta' JM\rangle$.

The above results lead to the following claim.

\medskip\noindent{\bf Claim 2:} Let $\{ Z^\dagger_{\mu\beta JM}(a^\dagger)|0\rangle
\}$ denote a complete set of orthonormal states, as defined in Claim 1, that
satisfy the constraint
$\hat S_-(uv) Z_{\mu\beta JM}^\dagger (a^\dagger)|0\rangle = 0$. 
Then the $N$-particle states number-projected from the corresponding 
quasi-particle states $\{  Z^\dagger_{\mu\beta JM}(\alpha^\dagger)|x\rangle \}$
form a (nonorthonormal) basis for the $N$-particle nucleus.
\medskip

\section{VCS wave functions of number-projected states}

The number-projected state $|n(x)\rangle$ has VCS wave function
\begin{equation} \Psi_{n(x)}(z) = \xi_0 \langle 0| e^{\hat S_-(z)} |n(x)\rangle = 
\xi_0  \Phi^s_n (zx) \,,\end{equation} 
where $\Phi^s_n= n! P_n\Phi^s$ and $P_n\Phi^s$ is the component of $\Phi^s$ of
degree
$n$ in its argument. It turns out, as shown in refs.\ \cite{RSC,CSR} that the
function $\Phi^s_n$ is  well-known in the theory of symmetric polynomials
\cite{Macd} as a Schur function (or S-function); such functions are the characters
of fully antisymmetric irreps of the unitary groups and easy to derive.

The definition 
\begin{equation} \Phi^s_n(zx) = \langle 0| e^{\hat S_-(z)} [\hat S_+(x)]^n |0\rangle \end{equation}
implies that
\begin{equation} \Phi^s_n(zx) = [\Gamma(\hat S_+(x)) \Phi^s_{n-1}](zx)   \,.\end{equation}
where  
\begin{equation} \Gamma(\hat S_+(x)) = \sum_j x_j \Gamma(\hat S^j_+))=
\sum_j x_jz_j\Big(2s_j - z_j {\partial\over \partial z_j}\big) \,.
\end{equation}
Thus, the needed Schur functions satisfy 
\begin{equation} \Phi^s_n(z) = \sum_j z_j \Big(2s_j - z_j {\partial\over \partial
z_j}\Big) \Phi^s_{n-1}(z)\,.\end{equation}

Following the methods of ref.\ \cite{RSC,CSR}, this recursion relation is solved by
making a change of variables from $\{ z_j\}$ to the symmetric power functions
\begin{equation} \phi^s_m = \sum_j 2s_j z_j^m \,.\end{equation}
The recursion relation then becomes
\begin{equation} \Phi^s_n = \Big[ \phi^s_1 - \sum_m^{n-1} m\, \phi^s_{m+1}\nabla^s_m \Big]
\Phi^s_{n-1} \,,\end{equation}
where
\begin{equation} \nabla^s_m = {\partial\over \partial \phi^s_m}\,,\end{equation}
and has solutions
\begin{equation} \begin{array}{ccl}  
\Phi^s_0 &=& 1 \\
\Phi^s_1 &=& \phi^s_1 \\
\Phi^s_2 &=& (\phi^s_1)^2-\phi^s_2 \\
\Phi^s_3 &=& (\phi^s_1)^3 - 3\phi^s_1\phi^s_2 + 2\phi^s_3 \,,\quad
{\rm etc.}\\
\end{array} \end{equation}
The general expression is obtained by means of the  identity
\begin{equation} \nabla^s_m \Phi^s_n = (-1)^{m+1} {1\over m} {n!\over (n-m)!} \Phi^s_{n-m} \,,
\label{eq:4.47}\end{equation}
derived in ref.\ \cite{RSC}.
With this identity, the recursion relation is given in the useful form
\begin{equation} \Phi^s_n = \sum_{m=1}^n (-1)^{m+1} { (n-1)!\over (n-m)!} \phi^s_m \Phi^s_{n-m}
\end{equation}
and has explicit solution given by
\begin{equation} \Phi^s_n = \det \left|\matrix{ \phi^s_1 & 1& 0&0 & \cdots & 0\cr
\phi^s_2 & \phi^s_1 &2&0 & \cdots & 0\cr
\phi^s_3 &\phi^s_2 & \phi^s_1 &3 & \cdots & 0\cr
&& \cdots &&\cr
\phi^s_n&\phi^s_{n-1} & \phi^s_{n-2} & \phi^s_{n-3} & \cdots & \phi^s_1 \cr}\right|
\,.\end{equation}
Some properties of these functions are discussed in refs.\
\cite{RSC,CSR,Macd}.

It follows from Claim 1 that the VCS wave function for a one quasi-particle state
$u_j\alpha^\dagger_{jm} |x\rangle$ has values given by 
\begin{equation} \Psi_{x1jm}(z) = \xi_{jm}\langle 0|a^{jm} e^{\hat S_-(z)} e^{\hat S_+(x)}
a^\dagger_{jm}|0\rangle \,.\end{equation}
From the observation that
\begin{eqnarray} \hat S_-^k a^\dagger_{jm} |0\rangle &=& 0 \quad \forall k \,, \\
\hat S_0^k a^\dagger_{jm} |0\rangle &=& ( -s_k +\half
\delta_{kj})a^\dagger_{jm}|0\rangle
\end{eqnarray}
it follows that
\begin{equation} \Psi_{x1jm}(z) = \xi_{jm}\Phi^j(zx) \,,\end{equation}
where
\begin{equation} \Phi^j(z) = (1+z_j)^{2s_j-1} \prod_{k\not= j} (1+z_k)^{2s_k} \,.\end{equation}
This function has values
\begin{equation} \Phi^j(z) = {1\over 1+z_j} \Phi^s(z)= \Phi^{s'}(z) \,,\end{equation}
where $s'$ is the set of quasi-spins with
\begin{equation} s'_k = s_k-\half \delta_{kj} \,.\end{equation}
Thus, by considering $\Phi^s$ to be one of a set of
functions parameterized by values $\{ s_j= (2j+1)/4\}$ of the quasispins,
$\Phi^j$ is seen to be the member of the set with $s_j$ replaced by $s_j-1/2$.
This substitution will be denoted by an operator $\Delta^j$, i.e.,  
\begin{equation} \Delta^j : s_j \to s_j-\half \,,\end{equation}
so that
\begin{equation} \Phi^j = \Delta^j \Phi^s \,.\end{equation}
The adjustment of the coherent state wave function $\Phi^s\to\Phi^j$ by the
replacement $s_j\to s_j-1/2$ to take account of the occupation of one state of a
single-particle level is an expression of the well-known {\em blocking
effect\/}.

The number-projected one quasi-particle state
\begin{equation} |n(x) 1jm\rangle =n! u_j\Pi_{2n+1} \alpha^\dagger_{jm} |x\rangle
=  [\hat S_+(x) ]^n a^\dagger_{jm} |0\rangle \end{equation}
is now seen to have VCS wave function with
\begin{equation} \Psi_{n(x) 1jm} (z) = n!\xi_{jm} P_n \Phi^j(zx) = \xi_{jm}
\Phi^j_n (zx)\,,\end{equation}
where $\Phi^j_n$ is the Schur function
\begin{equation} \Phi^j_n = \Delta^j \Phi^s_n \,.\end{equation}

Similarly, VCS wave functions are derived for all other number-projected
quasi-particle states.
For example, the seniority two
states projected from a two quasi-particle state 
$n! u_{j_1} u_{j_2} [\alpha^\dagger_{j_1} \otimes\alpha^\dagger_{j_2}]_{JM}
|x\rangle$ with  $J\not= 0$ have wave functions
\begin{equation} \Psi_{n(x) (j_1j_2)JM}(z) = [\xi_{j_1}\otimes \xi_{j_2}]_{JM}
\Phi^{j_1j_2}_n(zx) \,,\end{equation}
where 
\begin{equation} \Phi^{j_1j_2}_n = \Delta^{j_1}\Delta^{j_2}\Phi^s_n = \Phi^{s''}_n \end{equation}
is the Schur function for the quasi-spin set $s''$ with components
\begin{equation} s''_k = s_k - \half (\delta_{kj_1} + \delta_{kj_2}) \,.\end{equation}

The seniority-zero state $\hat S_+(y) |n(x)\rangle$ has VCS wave
function given by
\begin{equation} \xi_0 \sum_j y_jz_j\Big(2s_j - z_j {\partial\over \partial z_j}\big)
\Phi^s_n(zx) = \xi_0 \big[ \phi^s_1(\tilde y)  -
\sum_m m\phi_1(\tilde y\tilde x^m) \nabla_m ]\Phi^s_n(\tilde x) \,,
\end{equation}
where $\tilde x = xz$ and $\tilde y=yz$.
Thus, with the identity (\ref{eq:4.47}), the VCS wave function for the
state  $\hat S_+(y) |n(x)\rangle$ is given by
\begin{equation} \xi_0 \langle 0| e^{\hat S_-(z)}\hat S_+(y) |n(x)\rangle
=\xi_0\sum_{m=0}^n (-1)^m {n!\over (n-m)!} \phi_1(\tilde y\tilde x^m) \Phi_{n-m}
(\tilde x) \,. \label{eq:3.64}\end{equation}
As expected, this wave function becomes identical to $\xi_0 \Phi^s_{n+1}$
when $y=x$.

\section{Evaluation of energies and matrix elements}

To illustrate the techniques, we consider the simple BCS Hamiltonian
\begin{equation} H = \sum_j \varepsilon_j \hat n_j - G\hat S_+\hat S_- \,,\end{equation}
with
\begin{equation} \hat n_j = \sum_m a^\dagger_{jm} a^{jm} \,,\quad \hat S_\pm = \sum_j \hat
S^j_\pm \,.\end{equation}
For given real values of $x=(x_{j_1}, x_{j_2}, \ldots )$, the quasi-particle
operators are defined by
\begin{equation} \begin{array}{c}
\alpha^\dagger_{jm} = u_j a^\dagger_{jm} - v_j a_{jm} \,,\\
\alpha_{jm} = u_j a_{jm} + v_j a^\dagger_{jm} \,, \\
\end{array} \end{equation}
such that $u_j^2 + v_j^2 = 1$ and $v_j = x_j u_j$ so that
\begin{equation} u_j^2 = {1\over 1+x^2_j} \,,\quad u_jv_j = {x_j\over 1+x_j^2} \,,\quad v_j^2
= {x_j^2\over 1+x_j^2}\,.\end{equation}

\subsection{The number-projected vacuum energy}

From standard BCS theory, the expectation of $H$ for a quasi-particle vacuum
state is given by
\begin{equation} \langle x|H|x\rangle = \Big[ \sum_j 2s_j (2\varepsilon_j - Gv_j^2)v_j^2 - G
\Big( \sum_j 2s_j u_jv_j\Big)^2\,\Big] \Phi^s(x^2) \,.\end{equation}
With the above expressions for $u_j$ and $v_j$, this expression becomes
\begin{eqnarray} \langle x|H|x\rangle &=& \Big[ \sum_j 2\varepsilon_j{2s_j x^2_j\over
1+x^2_j}  -G\sum_j {2s_jx_j^4\over (1+x^2_j)^2} - G \Big( \sum_j {2s_j x_j\over
1+x^2_j}
\Big)^2\,\Big] \Phi^s(x^2) \nonumber\\
&=& \sum_j 4s_j \varepsilon_j x^2_j \Phi^j (x^2)
-G\sum_j 2s_jx_j^4 \Phi^{jj}(x^2) \nonumber\\
&&- G\sum_{ij} 4s_is_j x_ix_j \Phi^{ij}(x^2) 
\,.\end{eqnarray}
The energy of the $n$-pair state $|n(x)\rangle$,
\begin{equation} E_0^n(x) ={ \langle x |H|n(x)\rangle\over\langle x|n(x)\rangle}  
\,,\end{equation}
is now obtained by picking out the components of $\langle
x|H|x\rangle$ and $\langle x|x\rangle$ of degree $n$ in $x^2$.
Recalling that $P_{2n} \Phi^s = \Phi^s_n/n!$, we immediately obtain
\begin{eqnarray}  E_0^n(x) &=& {n\over \Phi^s_n(x^2)} \sum_j \Big[
4s_j\varepsilon_j x^2_j \Phi^j_{n-1} (x^2)
-G(n-1) 2s_jx_j^4 \Phi^{jj}_{n-2}(x^2) \nonumber\\
&& \qquad - G\sum_{i} 4s_is_j x_ix_j
\Phi^{ij}_{n-1}(x^2) \Big] \,. \label{eq:5.74})
\end{eqnarray}
Thus, the values  of the $x$ coefficients can be fixed such that the energy
$E_0^n=E_0^n(x)$ is minimized and the variational equation
\begin{equation} {\partial\over \partial z_j} \, \langle z|(H-E_0^n)|n(x)\rangle\Big|_{z=x}
=0\,, \quad \forall\, j\,,\label{eq:3.56} \end{equation}
is satisfied.

\subsection{Number-projected one quasi-particle energies}

As observed above, the $(2n+1)$-particle component of the one quasi-particle
state $n!u_j\alpha^\dagger_{jm}|x\rangle$ is the state $|n(x)1jm\rangle
=a^\dagger_{jm} |n(x)\rangle$. Thus, we consider the energies of number-projected
one quasi-particle states defined by
\begin{equation} E_{j}^n\, \langle n(x)1jm|n(x)1jm\rangle = \langle n(x)1jm| (H-E_0^n)
|n(x)1jm\rangle \end{equation} 
or equivalently by
\begin{equation}  E_j^n \, \Phi^j_n(x^2) = \langle x|a^{jm}( H-E_0^n)
a^\dagger_{jm}|n(x)\rangle \,.\end{equation} 
Now the  coherent state representation of the number operator
$\hat n_j = \sum_m a^\dagger_{jm} a^{jm}$, when acting on a seniority zero state,
is given by
\begin{equation} \Gamma (\hat n_j) \to 2 z_j {\partial\over \partial z_j} \,.\end{equation}
Therefore 
\begin{eqnarray} \langle x| a^{jm}a^\dagger_{jm}(H-E_0^n) |n(x)\rangle 
&=& \Big(1 -{z_j\over  2s_j} {\partial\over \partial z_j}\Big)\langle z|(H-E_0^n)
|n(x)\rangle\Big|_{z=x}\nonumber\\
&=& 0\end{eqnarray}
when $x$ is assigned the value for which eqn.\ (\ref{eq:3.56}) is satisfied.
It follows that
\begin{equation}  E^j_n \, \Phi^j_n(x^2) =  \langle x| a^{jm} [H,a^\dagger_{jm}] |n(x)\rangle
\,.\end{equation}

If we now (temporarily) regard $x$ as a variable parameter, we obtain from
Claim 1 the identity
\begin{eqnarray}  \langle x| a^{jm} [H,a^\dagger_{jm}] |n(x)\rangle &=&
 u_j \langle x| \alpha^{jm} [H,a^\dagger_{jm}] |n(x)\rangle \nonumber\\
&=& n! P_nu_j \langle x| \alpha^{jm} [H,a^\dagger_{jm}] |x\rangle \,,
\label{eq:64}\end{eqnarray}
where $P_n$  picks out the component of degree $n$ in
$x^2$ in what follows it.
Note that for a particular value of $x$, the matrix element 
$\langle x| \alpha^{jm} [H,\alpha^\dagger_{jm}] |x\rangle$
is a number.  However, by regarding $x$ first as a variable and evaluating the
matrix element as a function of $x$, it is meaningful to extract the component of
this function of degree $n$; the rhs of eqn.\ (\ref{eq:64}) is then the value of
this component at the assigned value of $x$.

The quasi-particle vacuum properties of the state $|x\rangle$,
which hold for arbitrary $x$, can now be used to put the matrix element on the
right into the standard equations-of-motion form  \cite{EofM}
\begin{equation}  \langle x| \alpha^{jm}[H,a^\dagger_{jm}] |x\rangle
=\langle x|\{ \alpha^{jm} ,[H,a^\dagger_{jm}] \} |x\rangle \,.\end{equation}
This matrix element is then evaluated by BCS methods to give
\begin{equation} u_j\langle x|\{ \alpha^{jm},[H,a^\dagger_{jm}]\} |x\rangle
= \big[ \varepsilon_j u_j^2 - G u_j^2 v_j^2 + Gu_jv_j \sum_i 2s_i u_iv_i\big]
\Phi^s(x^2) \,.
\end{equation}
The final step of number projection is now easy and gives the expression for the
number-projected one quasi-particle energies
\begin{equation} E_j^n \Phi^j_n(x^2) = \varepsilon_j \Phi^j_n(x^2)- nGx^2_j
\Phi^{jj}_{n-1}(x^2) + nG\sum_i 2s_i x_ix_j \Phi^{ij}_{n-1}(x^2) \,.\end{equation}
This is an explicit and simple expression for $E_j^n$ which involves only the
values of known (Schur) functions at the specific $x$ for which $E_0^n(x)$ is a
minimum.

\subsection{Matrix elements between seniority-two states}

Let $\{ A^\dagger_{klJM}(a^\dagger)|0\rangle\}$ be an orthonormal set of
two-particle seniority-two  states. 
The matrix elements between the corresponding
number-projected quasi-particles states are given by
\begin{eqnarray} &\langle x|A^{ijJM} HA^\dagger_{klJM} |n(x)\rangle 
= n!P_n  \langle x |A^{ijJM}(u\alpha)
H A^\dagger_{klJM}(u\alpha^\dagger) |x\rangle \phantom{xxxxxxx} & \nonumber\\
& =  n! P_n \Big(  \delta_{ij,kl} u_k^2u_l^2 \langle x|H|x\rangle + 
\langle x| [ A^{ijJM}(u\alpha),[
H, A^\dagger_{klJM}(u\alpha^\dagger)]] |x\rangle \Big) \,.&
\end{eqnarray}
Thus, the matrix elements for seniority-two
states are given by number projection of corresponding quasi-particle
Tamm--Dancoff expressions, as defined in the
equations-of-motion formalism \cite{EofM}. 
Moreover, the projections are
accomplished by expanding the unprojected expressions in terms of Schur functions.

\subsection{Matrix elements between seniority-zero states}

After evaluating the overlaps
$\langle x|\hat S_-(x) \hat S_+(y)|n(x)\rangle$ and $\langle x|\hat S_-(y) \hat
S_+(y')|n(x)\rangle$, one can construct an orthonormal set of seniority-zero
states for a $2n$-particle nucleus.
However, so that use can be made of Claim 1, it is
convenient to start with operators
$\{\hat S_+(y)\}$ that satisfy the condition
\begin{equation} \hat S_-(uv) \hat S_+(y) |0\rangle = 0 \,.\end{equation} 

The overlap $\langle x|\hat S_-(x)\hat S_+(y)|n(x)\rangle = (n+1)\langle x|\hat
S_+(y)|n(x)\rangle$ is obtained from the value   
$\xi_0 \langle x|\hat S_+(y)|n(x)\rangle$ at
$z=x$ of the coherent state wave function for the  state $ S_+(y)|n(x)\rangle$; 
an expression for this wave function is given by eqn.\ (\ref{eq:3.64}).  The
second overlap is given by
\begin{eqnarray} \langle x|\hat S_-(y) \hat S_+(y')|n(x)\rangle
&=&\sum_{j,m>0} \sum_{j',m'>0} y_j y'_{j'} \langle x|
a^{j\bar m} a^{jm} a^\dagger_{j'm'} a^\dagger_{j'\bar m'}|n(x)\rangle
\nonumber\\
&=& \sum_j 2s_jy_jy'_{j}\Phi^{jj}_n(x^2) \,.\end{eqnarray}

The matrix element $\langle x|\hat S_-(x)H\hat S_+(y)|n(x)\rangle = (n+1)\langle
x|H\hat S_+(y)|n(x)\rangle$ is evaluated starting from the observation that
\begin{eqnarray} \langle x|(H-E_0^n)\hat S_+(y)|n(x)\rangle &=&\langle n(x)|(H-E^0_n)\hat
S_+(y)|x\rangle \nonumber\\
&=& \langle x|\hat S_-(y) (H-E^0_n) |n(x)\rangle^* \end{eqnarray}
and that
\begin{equation} \langle z| \hat S_-(y) (H-E^n_0) |n(x)\rangle =
\sum_j y_j z_j \Big(2s_j - z_j{\partial\over \partial z_j}\Big)
 \langle z|(H-E^n_0) |n(x)\rangle 
\end{equation}
vanishes when $z=x$ because of eqn.\ (\ref{eq:3.56}).
It follows that
\begin{equation} \langle x|\hat S_-(x)H\hat S_+(y)|n(x)\rangle = (n+1) E_0^n 
\langle x|\hat S_+(y)|n(x)\rangle \,.\end{equation}

Writing
\begin{equation} \langle x| \hat S_-(y) H \hat S_+(y') |n(x)\rangle = 
n! P_n \langle x | \hat S_-(y) H\hat S_+(y') |x\rangle \,,
\end{equation}
and defining
\begin{equation} \tilde S_-(u^2y) = \sum_{j,m>0} u^2_jy_j a^{j\bar m} a^{jm} \,, \quad
\tilde S_+(u^2y') = \sum_{j,m>0} u^2_jy'_j a^\dagger_{jm} a^\dagger_{j\bar m}
\,,\end{equation} 
we can use Claim 1 to obtain
\begin{equation} \langle x| \hat S_-(y) H \hat S_+(y') |n(x)\rangle = 
n! P_n \langle x| \tilde S_-(u^2y) H \hat S_+(u^2y') |x\rangle \,.
\end{equation}
Then, with the identity
\begin{equation} \langle x| \tilde S_-(u^2y)\hat S_+(u^2y')H |x\rangle
=\sum_j 2s_j u^4_j y_jy'_j \langle x|H|x\rangle \,,
\end{equation}
we obtain
\begin{eqnarray} \langle x| \hat S_-(y) H \hat S_+(y') |n(x)\rangle &=& 
n! P_n \Big[ \sum_j 2s_j u^4_j y_jy'_j \langle x|H|x\rangle \nonumber\\
&&\quad + \langle x| [\tilde S_-(u^2y),[ H \hat S_+(u^2y')]] |x\rangle \Big] \,.
\end{eqnarray}

\section{Generalized seniority}

Talmi \cite{Talmi} has defined the sequence of $J=0$ states proportional to $\{
[\hat S_+(x)]^n |0\rangle ;$ $n = 1, 2, \ldots \}$ as states of
generalized seniority zero. Likewise, the sequences of $J\not= 0$ states   $\{
[\hat S_+(x)]^n |2 JM\rangle ; n = 1, 2, \ldots \}$, where $|2JM\rangle$ is a
two-particle state that is annihilated by all the quasispin lowering operators
(i.e., $\hat S^j_- |2JM\rangle$ $= 0,\; \forall j$), are defined to be states of
generalized seniority two. States of higher generalized seniority are similarly
defined. Thus, the concept of generalized seniority identifies subsets of states
of the corresponding seniority (more precisely summed multishell seniority $\nu =
\sum_j
\nu_j$). 
The generalized seniority zero states, for example, exclude states that are
generated by combinations of the $\{\hat S^j_+\}$ operators that are not simply
multiples of $\hat S_+(x)$.  Such states, sometimes described as broken $\hat
S_+(x)$ pair states \cite{Lorazo}, are also described as generalized seniority-two
or higher states, depending on how many
$\hat S_+(x)$ pairs are broken.
The problem is that, while one can define two-particle $J=0$ parent states $\{
|N=2,J=0\rangle\}$ as states
that are annihilated by the $\hat S_-(x)$ lowering operators, in parallel with
lowest-weight quasispin states, the sequences of states  $\{ [\hat S_+(x)]^n
|2,J=0\rangle ; n = 1, 2, \ldots \}$ are not orthogonal to the generalized
seniority-zero states. Consequently, as Talmi has emphasized, generalized
seniority does not define a complete orthogonal scheme.

Nevertheless, generalized seniority can be given a  
definition that does characterize a complete set of orthogonal states for each
nucleus separately.
For example, one can define a generalized seniority-two creation operator $\hat
S_+(y)$ for the $2n$-particle nucleus such that
\begin{equation} \langle x| \hat S_+(y) |n(x)\rangle =0\,. \label{eq:3.96}\end{equation}
With the identity
\begin{equation} \langle z | \hat S_+(p) [\hat S_+(x)]^n |0\rangle = n!P_n 
\langle z| \hat S_+(yp) |x\rangle = n! P_n \sum_j y_j  {\partial\over \partial
x_j} \Phi^s (zx) \,,\end{equation}
this equation reduces to the easy to solve equation for the $y_j$
coefficients 
\begin{equation} \sum_j 2s_j y_jx_j \Phi^j_n(x^2) =0 \,.
\end{equation}
States of higher generalized seniority may be defined similarly.

 Such a definition, has not been used to our knowledge probably because it
would appear to destroy the simplicity and elegance of the concept.
However, the facility to carry out number-projection analytically suggests that
this may no longer be a significant concern.

\section{Application to a two-level model}

To check the above methods, they have been applied to a simple two-level model
having Hamiltonian $ H = \sum_j \varepsilon_j\hat n_j -  G\hat S_+\hat S_- $ with
$\varepsilon_1=0$  and $\varepsilon_2=1$; the excitation energy
$\varepsilon_2-\varepsilon_1 =1$ then sets the unit of the energy scale. For
simplicity, we also set
$s_1=s_2=s$ and considered a model nucleus with $n=2s$ nucleons; this corresponds
to a situation in which, in the $G=0$ limit, the lower level is fully occupied and
the upper level empty.

For such a two-level model, we can set
\begin{equation} x_1=1\,,\quad x_2 = x\,,\end{equation}
because replacing these parameters values by the substitution $x_1\to k$,
$x_2\to kx$ would only change the overall norm of the state
\begin{equation} |n(x)\rangle = [\hat S^1_+ + x\hat S^2_+ ]^n |0\rangle \to k^n
[\hat S^1_+ + x\hat S^2_+ ]^n |0\rangle \,.\end{equation}
The energy $E^n_0(x)$ of the state, with $n=2s$, is then evaluated directly  from
eqn.~(\ref{eq:5.74}) and the value of $x$, for which it is a minimum, determined.

\begin{figure}[thp]
\epsfxsize=3.5in
\centerline{\epsfbox{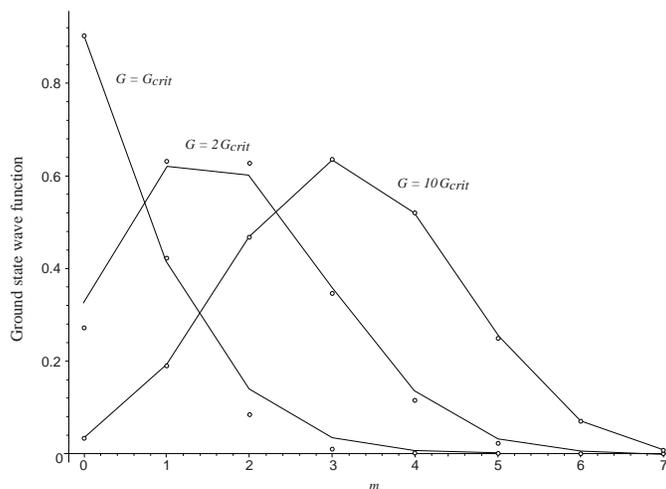}}
\caption{Ground state wave functions for $n=2s = 14$ particles in a two--level
pairing model for three values of the pairing interaction.  The expansion
coefficients, as defined in the text, for normalized number--projected states are
shown as small circles. The corresponding coefficients for exactly computed ground
states are connected by continuous lines. }
\label{fig:wfns}
\end{figure}

The structure of the state $|n(x)\rangle$ is revealed by expanding it on a basis of
states  labeled by the two-level quasispin quantum numbers
\begin{equation} \{ |m\rangle = |s,s-m; s,m-s\rangle\,,\quad m = -s, \ldots, +s\} \,.\end{equation}
The state $|m\rangle$ is the state with m nucleon pairs in
the upper level and $s-m$ pairs (or $m$ hole pairs) in the lower level.
The expansion 
\begin{equation} |n(x)\rangle = \sum_{m=-s}^s C_m |m\rangle \end{equation}
in this basis is easily inferred from the identity
\begin{equation} \Phi^s_{2s} (x^2) = \langle x|n(x)\rangle \,,
\end{equation}
which implies that the squares of the $C_m$ coefficients  
are  given by the corresponding expansion of $\Phi^s_{2s}(x^2)$ as a
polynomial in $x^{2m}$;
\begin{equation} \Phi^s_{2s}(x^2) = {1\over n!}\sum_m |C_m|^2  x^{2m} \,.\end{equation}
The $C_m$ coefficients are shown, for $s=7/2$ and three values of $G$, in
comparison with exactly computed wave functions for the model in
fig.~\ref{fig:wfns}.

For $s=7/2$, the model corresponds to 14 nucleons in two $j=13/2$ levels.
Results computed for other values of $n=2s$ show that the number-projected wave
functions rapidly become exact as $s\to\infty$.

As an indication of the improvement given by number projection over unprojected
BCS results is given by comparing the ratio of the number of nucleons occupying the
upper level to the number in the lower level.
The mean number in the upper level is given for the above wave functions by
\begin{equation} \langle n_2\rangle = 
\sum_m 2m |C_m|^2 x^{2m} \over \sum_m  |C_m|^2 x^{2m} \end{equation}
and the mean number for the lower level is 
\begin{equation} \langle n_1\rangle = 2s - \langle n_2\rangle \,.\end{equation}
Fig.~\ref{fig:ratios} shows the ratio $\langle n_2\rangle / \langle n_1\rangle$
for several values of the pairing interaction in comparison to the exact ratios
and those of the BCS approximation.
\begin{figure}[thp]
\epsfxsize=3.5in
\centerline{\epsfbox{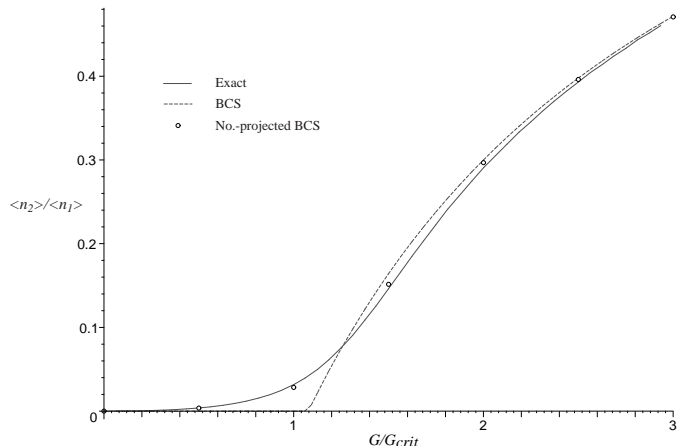}}
\caption{The ratio of the mean number of particles in the upper level to the
mean number in the lower level for the $n=2s=14$ two--level pairing model defined
in the text. The ratios for number--projected states are compared with exactly
computed results and with those of the BCS approximation.}
\label{fig:numratios}
\end{figure}
As $n=2s \to\infty$, the occupancies given by the BCS approximation are
found to approach those of an exact calculation.
However, for a finite number of particles, a considerable improvement is
gained by number projection (as expected from previous results).

\section{Concluding remarks}

The calculations that have been done, cf.\ for example refs.\ \cite{CSR,BPA},
show clearly that number projection effects very substantial improvement to the
accuracy of the BCS approximation in applications to finite nuclei. 
Moreover, the coherent state techniques introduced in refs.\ \cite{RSC} and
\cite{CSR} were found to facilitate the practical steps of carrying out the number
projection considerably.
These techniques have been developed in this paper in the hope 
that they will be taken up by others in calculations of the low-energy spectra of
nuclei in which pair-correlations are dominant.
Of particular interest are the low-energy spectra of single-shell nuclei with
both pairing and quadrupole interactions where the objective will be explore the
emergence of deformed rotational states.

Other recent developments that may be usefully deployed in concert with number
projection are:  the identification of a range of seniority-conserving
interactions \cite{CenConserv}, and analytic techniques to carry out angular
momentum projection from a class of deformed intrinsic states \cite{AMproj}.


\begin{references}

\bibitem{RSC} D.J.\ Rowe, T.\ Song, and H.\ Chen, Phys.\ Rev.\ C {\bf 44}
(1991) R598.

\bibitem{CSR}  H.\ Chen, T.\ Song, and D.J.\ Rowe, Nucl.\ Phys.\ A {\bf
582} (1995) 181.

\bibitem{Mottelson} B.R.\ Mottelson, in {\it The many--body problem\/}, (Dunod,
Les Houches, 1958).

\bibitem{DMP}  K.\ Dietrich, H.J.\ Mang, and J.H.\ Pradal, Phys.\ Rev.\
{\bf 135} (1964) B22.

\bibitem{projqp}  A.\ Lande, Ann.\ Phys.\ {\bf 31} (1965) 525; P.L.
Ottaviani and M. Savoia, Phys.\ Rev.\ {\bf 187} (1969) 1306; Nuovo
Cim.\ {\bf 67}A (1970) 630.

\bibitem{antigem}  A.J.\ Coleman, J.\ Math.\ Phys.\ {\bf 6} (1965) 1425;
J.V.\ Ortiz, B.\ Weaver, and Y.\"Ohrn, Int.\ J.\ Quantum Chem.\ Symp.\
{\bf 15} (1981) 113.

\bibitem{Lorazo} B.\ Lorazo, Phys.\ Lett.\ B {\bf 29} (1969) 150; Nucl.\
Phys.\ A {\bf 153} (1970) 255.

\bibitem{Gambhir} Y.K.\ Gambhir, A.\ Rimini, and T.\ Weber,  Phys.\ Rev.\
{\bf 188} (1969) 1573;  Phys.\ Rev.\ C {\bf 7} (1973) 1454.

\bibitem{Talmi} I.\ Talmi, Nucl.\ Phys.\ A {\bf 172} (1971) 1;
Phys.\ Lett.\ {\bf 55}B (1975) 255.

\bibitem{VP} J.P.\ Vary and A.\ Plastino,  Phys.\ Rev.\ C {\bf 28} (1983)
2494.

\bibitem{BPA}  K.\ Allaart, E.\ Boeker, G.\ Bonsignori, M.\ Savoia, and
Y.K.\ Gambhir, Phys.\ Reports {\bf 169} (1988) 209.

\bibitem{AMproj}  D.J.\ Rowe, S.T. Bartlett, and C.\ Bahri, Phys.\ Lett.\ 
 B {\bf 472}, 227-231 (2000); R.M.\ Asherova. Yu.F.\ Smirnov, V.N.\ Tolstoy, and
A.P.\ Shustov, Nucl.\ Phys.\ {\bf A355}, 25 (1981).

\bibitem{VCS} D.J.~Rowe, J.\ Math.\  Phys. {\bf 25}, 2662 (1984);
  D.J.~Rowe, G.~Rosensteel and R.~Carr,   J.\ Phys.\ A:
  Math.\ Gen. {\bf 17}, L399 (1984); D.J.~Rowe, G.~Rosensteel and 
  R.~Gilmore,  J.\ Math.\ Phys. {\bf 26}, 2787 (1985).

\bibitem{RR91} D.J.~Rowe and J.~Repka,   J.\ Math.\ Phys. {\bf 32}, 2614
  (1991).

\bibitem{Macd}  I.G.\ Macdonald, {\it Symmetric functions and Hall polynomials\/}
(Oxford, 1979).

\bibitem{EofM} D.J.\ Rowe, Rev.\ Mod.\ Phys.\  {\bf 40} (1968) 153; D.J.\ Rowe,
``Nuclear Collective Motion; Models and Theory" (Methuen, London, 1970).

\bibitem{CenConserv}  D.J.\ Rowe, ``Partially solvable pair--coupling models and
some challenges" (University of Toronto preprint).

\end{references}
\end{document}